\documentclass[12pt]{article}
\renewcommand{\baselinestretch}{1.0}
\newcommand{\be}{\begin{equation}}
\newcommand{\ee}{\end{equation}}
\newcommand{\br}{\begin{eqnarray}}
\newcommand{\er}{\end{eqnarray}}
\newcommand{\ra}{\rangle}
\newcommand{\la}{\langle}
\usepackage{epsf}
\usepackage{epsfig}
\begin{document}
\begin{center}
{\bf\Large Three Lectures on Chiral Symmetry\\
by\\
\vskip2.0cm
N.D. Hari Dass\\
Institute of Mathematical Sciences, Chennai 600 113, India}
\end{center}
\newpage
\section{Abstracts of Lectures}
\subsection{Lecture I}
In this lecture I shall begin by tracing the physical origins of chiral
symmetry. By asking for the conditions for the conservation of the {\em
Axial Vector Current} one arrives at the {\it Goldberger Triemann Condition}
and the need for a triplet of massless scalars to be identified with {\em Pions}. The concept of {\it Partially Conserved Axial Current} or PCAC is shown to emerge naturally for the real world pions which are not massless. I shall work out the full group structure of chiral symmetry. The {\it Linear
Sigma Model} is shown to be the simplest realisation.
\subsection{Lecture II}
In this lecture I shall show how one needs {\em Spontaneous breakdown of
chiral symmetry} to make the Sigma-model to be compatible with nucleon mass.
The phenomenon of spontaneous breaking of continuous symmetries will be
elaborated with the example of $O(4)$ group. In this lecture I shall also
discuss the notion of {\em Nonlinear Realisation} of chiral symmetry.
\subsection{Lecture III}
In this lecture I shall discuss the dramatic effects brought about by 
fluctuations in the Goldstone Boson fields. I shall show that in the
large-N limit of $O(N)$-models in four dimensions the fluctuations completely
destroy the spontaneously broken phase. 

If time permits, I shall briefly discuss the implications for nuclear forces and
I shall conclude my talks with a {\em Renormalisation Group Overview} and with a proposal for how to do {\em Chiral Perturbation Theory} better.
\newpage
\section{Lecture I: Why Chiral Symmetry?}
Let us begin by first reviewing {\em Noether's Theorem} which is central
to any discussion of {\it Symmetries and Conservation Laws}. Consider the
{\em action functional}
\be
S = \int d^d x {\cal L}(\psi,\partial\psi,..)
\ee
where $\psi$ stands collectively for all the relevant fields. Let the action
be invariant under the {\it global} transformations
\be
\label{trans}
\delta\psi = i\Lambda \psi
\ee
Now let us consider the variation of the action under eqn(\ref{trans}) but
now for a space-time {\em dependent} $\Lambda(x)$. The action is {\bf not
invariant} now but the variation of the action must depend {\bf only} on the {\em derivatives of $\Lambda(x)$}. For a generic $\Lambda(x)$ this variation will be
proportional to $\partial_\mu \Lambda(x)$:
\be
\label{noethervar}
\delta S = -\int \partial_\mu \Lambda(x)\cdot j^\mu(x)
\ee
Now we consider the {\em variational principle} determining the equations
of motion of the field theory. Here one considers variations $\tilde\delta$
of fields that vanish on some {\em initial} and {\em final} configurations
that leaves the {\em extremises the action}. In these situations
\be
\label{boundvar}
{\tilde\delta}S = \int {{\tilde\delta}S\over {{\tilde\delta}\psi}}{\tilde\delta}\psi
\ee
We are looking for a {\em specific} field configuration, usually called $\psi_{cl}$ for which ${\tilde\delta S} = 0$:
\be
{{\tilde\delta S\over {\tilde\delta}\psi}}|_{\psi=\psi_{cl}}=0
\ee
Returning to the context of Noethers theorem the variation of the action
eqn(\ref{noethervar}) can be rewritten as
\be
\delta S = \int \Lambda(x)\partial_\mu j^\mu
\ee
where we have tacitly assumed that fields off sufficiently fast at 'infinity'.
This is equivalent to demanding vanishing fields on a boundary and we can use
eqn(\ref{boundvar}) along with eqn(\ref{trans}) for {\em space-time dependent}
$\Lambda(x)$ to get
\be
\delta S = \int {{\delta S\over \delta\psi}}\Lambda(x)~\psi
\ee
finally arriving at the {\em Noether Theorem}
\be
\partial_\mu j^\mu = {\delta S\over \delta\psi}~\psi
\ee
In particular, if the fields $\psi$ satisfy the equations of motion, the {\it currents} $j_\mu(x)$ are {\em conserved}:
\be
\partial_\mu j^\mu = 0
\ee
Let us apply this to the case of Dirac action of the form
\be
S_{gen} = \int d^dx {\bar\psi}(x) (i{\gamma^\mu~\partial_\mu - M}) \psi(x)
\ee
Let us consider the transformation law
\be
\delta\psi = i\lambda \psi
\ee
when $\lambda$ is a {\em real constant} the action is obviously {\em invariant}. When
$\lambda$ is position dependent one gets
\be
\delta S = -\int {\bar\psi}(x) \gamma^\mu \psi(x) \partial_\mu\lambda(x)
\ee
We identify the conserved current to be
\be
j^\mu(x) = \bar\psi(x)\gamma^\mu\psi(x)
\ee
{\bf Exercise}\\
Show that this current is conserved only when $\psi$ satisfies the equation
of motion.

Now let us turn to the case where $\psi$ includes protons $\psi_p$ and neutrons $\psi_n$. The electromagnetic current of the protons
\be
j^\mu_{em}=\bar\psi_p(x)\gamma^\mu\psi_p(x)
\ee
is conserved
\br
i\partial_\mu j^\mu_{em}& = & {\bar\psi}_p(x)\gamma^\mu\partial_\mu\psi_p(x)
+(\partial_\mu\bar\psi_p(x))\gamma^\mu\psi_p\nonumber\\
& = & {\bar\psi}_p M_p \psi_p+(-M_p){\bar\psi}_p\psi_p = 0
\er
In contrast the 'current' ${\bar\psi}_p\gamma^\mu\psi_n$ is not conserved but is instead equal to
\be
i\partial_\mu({\bar\psi}_p\gamma^\mu\psi_n) = (M_p-M_n){\bar\psi}_p\psi_n
\ee
However, this current is also conserved if $M_p=M_n$ which obviously is a
situation of greater symmetry. In fact the {\it vector current} ${\vec J}^\mu =\bar\psi \gamma^\mu{\vec\tau}\psi$(now $\psi$ is an isospinor-spinor) is conserved then
\be
\partial_\mu {\vec J}^\mu = 0
\ee
This enhanced symmetry is called {\em Isospin invariance}.
\subsection{Axial Vector Currents}
Now we can ask whether the {\it axial vector current} ${\vec J}_5^{(p)\mu} = {\bar\psi}_p\gamma^\mu\gamma_5\psi_p$ can likewise be made to be conserved with
special choice of parameters. It is easily checked that
\be
i\partial_\mu \bar\psi_i~\gamma^\mu~\gamma_5\psi_j = -(M_i+M_j)\bar\psi_i\gamma_5\psi_j
\ee
Since the {\it mass matrix} ${\cal M}$ whose eigenvalues are $M_i$(i=1,2
corresponds to protons and neutrons) is {\it positive}, it follows that
no choice of ${\cal M}$ will lead to a {\em conserved axial current}, 
{\bf unless all the masses are zero!} This latter possibility is not
as unphysical as it may sound!!

One can ask whether at all axial vector current could ever be conserved. 
Considering the symmetry with which {\em axial} and {\em vector} currents
occur in the {\it Weak Interaction Hamiltonian} it would indeed be very
unusual if there was a such a disparity between them.

\begin{figure}[htb]
\begin{center}
\mbox{\epsfig{file=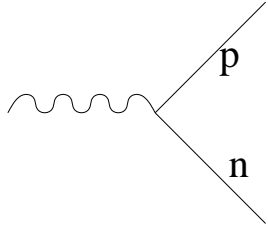,height=3truecm,angle=0}}
\end{center}
\caption{Direct contribution}
\label{fig:direct}
\end{figure}

Let us probe this further by concentrating on the {\em matrix elements}
of the axial vector current between one proton and one neutron states.
The {\it direct contribution} corresponding to the fig(\ref{fig:direct})
is given by
\be
\label{direct}
\la p|A_\mu^{(+)}|n \ra_{dir} = G_A \bar u(k_p)\gamma_\mu \gamma_5 u(k_n)
\ee
Here the superscript $(+)$ on the axial current refers to the component
that transfers one unit of charge i.e it connects ingoing neutron states
to outgoing proton states. As discussed above this part of the axial vector is
{\bf not conserved}.

\begin{figure}[htb!]
\begin{center}
\mbox{\epsfig{file=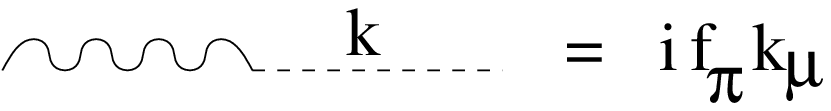,height=1truecm,angle=0}}
\end{center}
\caption{Coupling to Pseudoscalar }
\label{fig:coupling}
\end{figure}

Suppose there existed a {\it pseudoscalar positively charged particle} $\phi^{(+)}$ which
{\bf interacts with nucleons} and whose axial current has a contribution
of the form
\be
A_\mu^{(+)} = f_\phi \partial_\mu \phi^{(+)}+...
\ee
Diagrammatically this can be represented by fig(\ref{fig:coupling}).

\begin{figure}[htb]
\begin{center}
\mbox{\epsfig{file=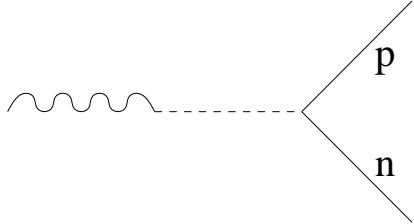,height=3truecm,angle=0}}
\end{center}
\caption{Pseudoscalar mediated contribution}
\label{fig:mediated}
\end{figure}

This gives rise to an additional contribution to the matrix element which is
diagrammatically shown in fig(\ref{fig:mediated}) given by
\be
\label{mediated}
\la p|A_\mu^{(+)}|n \ra_{med} = ig_{NN\pi}{if_\phi k_\mu\over k^2-m_\phi^2} \bar u(k_p)\gamma_5 u(k_n)
\ee
Thus the total contribution to the single-nucleon matrix elements of the axial
current is given by
\be
\la p|A_\mu^{(+)}|n\ra_{tot} = G_A\bar u(k_p)\gamma^\mu\gamma_5 u(k_n)-{f_\phi g_{NN\pi}k^\mu\over (k^2-m_\phi^2)}\bar u(k_p)\gamma_5 u(k_n)
\ee
It is then easy to verify that
\be
\la p|\partial_\mu A^{(+)\mu}|n\ra = [2G_AM_N - f_\phi g_{NN\pi}{k^2\over(k^2-m_\phi^2)}]\bar u(k_p)\gamma_5u(k_n)
\ee
This is zero if $m_\phi^2=0$ {\bf and} 
\be
\label{Goldberger-Trieman}
2G_AM_N=f_\phi g_{NN\phi} 
\ee
In nature
there is indeed the {\em pion} satisfying the properties conjectured here for
$\phi$. Thus the principle of a conserved axial vector current could have
actually predicted the existence of pions!

The new symmetry associated with the conservation of axial vector currents
is called {\bf Chiral Symmetry}.

In eqn(\ref{Goldberger-Trieman}), called the {\em Goldberger-Trieman Relation}, $f_\pi$ can be determined from pion decay and
all the other quantities are also determined from direct observations. That
the observed quantities satisfy this relation to very good accuracy is a
strong evidence in favour of the ideas expounded here.

But in reality $m_\pi^2 \ne 0$ though it is very small compared to $M_N^2$.
Keeping this in mind, one has
\be
\label{pcac1}
\la p|\partial_\mu A^{(+)\mu}|n\ra = f_\pi g_{NN\pi} m_\pi^2 {1\over k^2-m_\pi^2}\bar u(k_p)\gamma_5 u(k_n)
\ee
In the above-mentioned considerations the coupling between pions and nucleons
has been described by
\be
{\cal L}_{NN\pi} = g_{NN\pi}\bar\psi_p \gamma_5 \pi^- \psi_n
\ee
Completing the argument for all the three components of the axial vector current
means invoking a triplet of pions forming an iso-vector and the full 
pion-nucleon coupling takes the form
\be
\label{lagint}
{\cal L}_{NN\pi} = g_{NN\pi}\bar\psi_p {\vec\tau}\cdot{\vec\pi}\gamma_5\psi_n
\ee
The pions can at this stage be described by
\be
\label{lagpi}
{\cal L}_\pi = {1\over 2}[(\partial_\mu\vec \pi)^2-m_\pi^2 \pi^2]
\ee
The generalisation of eqn(\ref{pcac1}) to the triplet of pions is
\be
\label{pcac2}
\la N,p|\partial_\mu {\vec A}^{\mu}|N,q\ra = f_\pi g_{NN\pi} m_\pi^2 {1\over k^2-m_\pi^2}\bar U(p)\gamma_5{\vec \tau} U(q)
\ee
The Klein-Gordon equation for pions resulting from eqns(\ref{lagpi},\ref{lagint}) is
\be
(\partial_\mu\partial^\mu-m_\pi^2)\vec\pi = g_{NN\pi}\bar N \gamma_5 \vec\tau N
\ee
One sees that {\it atleast as far as the single nucleon matrix elements are
concerned} eqn(\ref{pcac2}) can be reproduced by
\be
\partial_\mu {\vec A}^\mu = f_\pi m_\pi^2 \vec\pi
\ee
This equation called {\em Partially Conserved Axial Current} is valid more
generally than what is indicated here. Let us also record the equation of
motion for the nucleon fields arising out of
\be
{\cal L}_N = \bar N (i\gamma^\mu\partial_\mu-M) N
\ee
and eqn(\ref{lagint})
\be
(i\gamma^\mu\partial_\mu-M)N = g_{NN\pi}\gamma_5{\vec \tau}\cdot{\vec\pi}N
\ee
The candidate expression for the (conserved) axial-vector vector current is
\be
{\vec A}^\mu = \bar N \gamma^\mu\gamma_5{\vec\tau\over 2}N + f_\pi\partial^\mu{\vec\pi}
\ee
The form of the pionic contribution to the axial current is strongly supported
from Pion weak decays(in fact $f_\pi$ was introduced there)
\newpage
\section{Lecture II: More on axial vector currents}
If one starts from the pion-nucleon Lagrangean introduced so far
\be
{\cal L}_{\pi N} = {1\over 2}[(\partial_\mu\vec\pi)^2+m_\pi^2{\vec\pi}^2]+
\bar N(i\gamma^\mu\partial_\mu-M)N -ig_{NN\pi}\bar N \gamma_5\vec\tau\cdot\vec\pi N
\ee
If one makes the natural generalisation of the electromagnetic gauge transformation to
\be
\delta_V \psi = i\vec\Lambda\cdot{\vec\tau\over 2}\psi
\ee
one can see that the pion-nucleon lagrangean is invariant if the pions transform according to
\be 
\delta_V \vec\pi = -\vec\Lambda\times\vec\pi
\ee
when $\Lambda$ is constant. If however $\Lambda$ is {\em position dependent}
one finds the variation of the Lagrangean to be
$$
-\partial_\mu\vec\Lambda\{-\partial_\mu\vec\pi\times\vec\pi+\bar N\gamma^\mu{\vec\tau\over 2}N\}
$$
leading to the isovector vector current
\be
{\vec V}_\mu = \bar N\gamma_\mu{\vec\tau\over 2}N-\partial_\mu\vec\pi\times\vec\pi
\ee
The equations of motion resulting from the above lagrangean are
\br
(i\gamma^\mu\partial_\mu-M) N &=& ig_{NN\pi}\gamma_5 \vec\tau\cdot\vec\pi N\nonumber\\
\bar N (-i\gamma^\mu\partial_\mu-M) &=& ig_{NN\pi}\bar N\gamma_5\vec\tau\cdot\vec\pi\nonumber\\
(\partial_\mu\partial^\mu-m_\pi^2)\vec\pi &=& -ig_{NN\pi}\bar N \gamma_5 \vec\tau N
\er
Using these it is easy to show that
\be
\partial^\mu {\vec V}_\mu = 0
\ee
Now let us see how far we can mimic this procedure for the axial vector currents. We can introduce the axial transformations
\be
\delta_A \psi = i\vec\omega{\vec\tau\over 2}\gamma_5 \psi;~~~~~\delta_A\vec\pi = ?
\ee
It follows that
\be
\delta_A \bar \psi = \bar\psi\gamma_5 i\vec\omega\cdot{\vec\tau\over 2}
\ee
The variation of the lagrangean {\bf omitting the purely pionic parts} is
\be
\delta_A {\cal L}^{-\vec\pi} = -iM\bar N \gamma_5 \vec\tau\cdot\vec\omega N
-ig_{NN\pi}\bar N\gamma_5\vec\tau\cdot\delta_A\vec\pi N + g_{NN\pi}\bar N\vec\pi\cdot\vec\omega N -\bar N \gamma^\mu\gamma_5{\vec\tau\over 2}\cdot\partial_\mu\vec\omega
\ee
One sees that by virtue of the Goldberger-trieman relation, the first two terms
would cancel each other if we choose
\be
\label{pitrans1}
\delta_A\vec\pi = -f_\pi\vec\omega
\ee
Under this the variation of the pionic lagrangean is
\be
\delta_A {\cal L}_{\vec\pi} = f_\pi m_\pi^2 \vec\omega\cdot\vec\pi - f_\pi\partial_\mu\vec\pi\cdot\partial^\mu\vec\omega
\ee
Thus the Noether prescription would point to an axial current of the form
\be
{\vec A}_\mu = \bar N \gamma^\mu \gamma_5{\vec\tau\over 2}N+f_\pi\partial_\mu\vec\pi
\ee
which is {\bf not exactly conserved} because of terms in the variation still
proportional to $\vec\omega$. While the term proportional to $m_\pi^2$ is
what we would have anticipated on the basis of PCAC there is an additional
term in $\delta_A {\cal L}$ equal to $g_{NN\pi}\bar N \vec\pi\cdot\vec\omega N$
which is clearly {\bf undesirable}.

To get a clue as to what is happening note
\be
\delta_V (i\bar N \gamma_5 \vec\tau N) = -\vec\Lambda\times(i\bar N \gamma_5
\vec\tau N)
\ee
Thus under vector transformations $(i\bar N \gamma_5\vec\tau N)$ transforms
{\em exactly} as $\vec\pi$. Let us see how this transforms under the 
{\bf axial} transformations:
\be
\delta_A(i\bar N \gamma_5\vec\tau N) = -\bar N N \vec\omega
\ee
But {\bf this does not look like} $\delta_A \vec\pi = -f_\pi \vec\omega$
that we proposed earlier!

But note that
\be
\delta_A \bar N N = \vec\omega\cdot(i\bar N \gamma_5\vec\tau N)
\ee
This {\em suggests} postulating a new field $\sigma$ that, along with $\vec\pi$,forms an {\em irreducible representation} of the {\em Chiral Symmetry Group}!
The corresponding transformation laws are
\br
\label{fulltrans}
\delta_V \vec\pi &=& -\vec\Lambda\times\vec\pi;~~~~\delta_A \vec\pi = -\sigma\vec\omega\nonumber\\
\delta_V \sigma &=& 0;~~~~~\delta_A \sigma = \vec\omega\cdot\vec\pi\nonumber\\
\delta_V N &=& i\vec\Lambda\cdot{\vec\tau\over 2}N;~~~~\delta_A N = i\vec\omega\cdot{\vec\tau\over 2}\gamma_5 N
\er
\subsection{Group Properties}
The way to compute the composition law for various transformations is to
look at
\be
\delta_{[1,2]} = [\delta_1,\delta_2]
\ee
Introducing the total variation of any field $\psi$ as
\be
\delta \psi = \delta_V \psi + \delta_A \psi
\ee
and the {\em generators} of vector transformations $\vec T$ and of axial
transformations $\vec X$ we can write
\be
\delta\psi = [\vec\Lambda\cdot\vec T +\vec\omega\cdot\vec X, \psi]
\ee
Then it is easily shown that
\br
\delta_{12}\sigma &=& \vec\omega_{12}\cdot\vec\pi = -\vec\pi\cdot[\vec\omega_2\times\vec\Lambda_1-\vec\omega_1\times\vec\Lambda_2]\nonumber\\
\delta_{12}\vec\pi &=& -(\vec\omega_1\times\vec\omega_2)\times\vec\pi
-(\vec\Lambda_1\times\vec\Lambda_2)\times\vec\pi -(\vec\Lambda_1\times\vec\omega_2-\vec\Lambda_2\times\vec\omega_1)\sigma
\er
Leading to the Lie algebra
\be
[T_i,T_j]=i\epsilon_{ijk}T_k;[T_i,X_j]=i\epsilon_{ijk}X_k;[X_i,X_j]=i\epsilon_{ijk}T_k
\ee
\subsection{Invariants}
It is easy to verify that the following are invariants:
\br
I_1 &=& \sigma^2+\vec\pi^2\nonumber\\
I_2 &=& (\partial_\mu \sigma)^2+(\partial_\mu \vec\pi)^2\nonumber\\
I_3 &=& \bar N i\gamma^\mu\partial_\mu N\nonumber\\
I_4 &=& \bar N N \sigma + i\bar N \gamma_5 \vec\tau N\cdot\vec\pi
\er
With these invariants we can build the following {\em invariant lagrangean}
which is bound to yield {\em conserved} axial and vector currents.
\be
\label{fulllag}
{\cal L}_{inv} = \bar N i\gamma^\mu\partial_\mu N - g_{NN\pi}\bar N(\sigma+i\gamma_5\vec\tau\cdot\vec\pi) N+{1\over 2}[(\partial_\mu\sigma)^2+(\partial_\mu\vec\pi)^2]-{\mu^2\over 2}(\sigma^2+\vec\pi^2)+{\lambda\over 4!}(\sigma^2+\vec\pi^2)^2
\ee
Let us illustrate Noether procedure by explicitly working out the conserved
axial and vector currents coming only from the $(\sigma,\vec\pi)$ part of
the invariant lagrangean. Remember we should work out the variation of
the lagrangean under {\em position dependent} $(\vec\Lambda(x),\vec\omega(x)$:
\br
\delta_A \{{1\over 2}[(\partial_\mu\sigma)^2+(\partial_\mu\vec\pi)^2]\} &=&
-\partial_\mu\vec\omega(\sigma\partial_\mu\vec\pi-\vec\pi\partial_\mu\sigma)\nonumber\\
\delta_V \{{1\over 2}[(\partial_\mu\sigma)^2+(\partial_\mu\vec\pi)^2]\} &=&
-\partial_\mu\vec\Lambda(\vec\pi\times\partial_\mu\vec\pi)
\er
giving
\br
{\vec A}_\mu &=& \sigma\partial_\mu\vec\pi - \vec\pi\partial_\mu\sigma
\nonumber\\
{\vec V}_\mu &=& \vec\pi\times\partial_\mu\vec\pi
\er
Adding the nucleon contribution, the full currents are
\br
\label{currents}
{\vec A}_\mu &=&\bar N \gamma_\mu\gamma_5{\vec\tau\over 2}N+ \sigma\partial_\mu\vec\pi - \vec\pi\partial_\mu\sigma
\nonumber\\
{\vec V}_\mu &=& \bar N \gamma_\mu{\vec\tau\over 2}N+\vec\pi\times\partial_\mu\vec\pi
\er
The full set of equations of motion are
\br
\label{fulleqn}
\partial_\mu\partial^\mu\vec\pi+\mu^2\vec\pi+{\lambda\over 6}(\sigma^2+\vec\pi^2)\vec\pi+ig_{NN\pi}\bar N\gamma_5\vec\tau N&=&0\nonumber\\
\partial_\mu\partial^\mu\sigma+\mu^2\sigma+{\lambda\over 6}(\sigma^2+\vec\pi^2)\sigma+g_{NN\pi}\bar N N&=&0\nonumber\\
(i\gamma^\mu\partial_\mu-ig_{NN\pi}\gamma_5\vec\tau\cdot\vec\pi-g_{NN\pi}\sigma)N&=&0
\er
It is easy to verify that the currents in eqn(\ref{currents}) are {\em exactly
conserved} when eqn(\ref{fulleqn}) are used.
\subsection{Spontaneous breaking of chiral symmetry}
While the fully invariant lagrangean eqn(\ref{fulllag}) gives an axial current
that is exactly conserved we note that eqn(\ref{fulllag}) has the following
{\bf severe shortcomings}:

i) {\em There is no nucleon mass term in the lagrangean}. As it stands it
describes {\bf massless} nucleons. However, there is an additional term
$g_{NN\pi}\bar N N$.

ii) {\em The masses of the pions and sigma are equal.} Phenomenologically
no isoscalar with such a property exists.

iii){\em The axial current is conserved even with nonvanishing pion mass!}

iv) {\em The axial current does not have a $f_\pi\partial_\mu\vec\pi$ term}.
However, there is a $\sigma\partial_\mu\vec\pi$ term.

v) {\em The pion transformation in eqn(\ref{fulltrans}) is $\delta_A\vec\pi = -\sigma\vec\omega$ while what we had sought in eqn(\ref{pitrans1}) was $\delta_A\vec\pi = -f_\pi\vec\omega$}.

It is remarkable that the difficulties i),iv) and v) can be simultaneously
resolved if we let $\sigma = f_\pi$! On the other hand that would break
chiral symmetry explicitly. What is really needed to fix the problem is
something like $\sigma = f_\pi+some fields$.

Let us consider the equations of motion only in the $(\sigma,\vec\pi)$ sector:
\br
\label{pieqn}
\partial_\mu\partial^\mu\vec\pi+\mu^2\vec\pi+{\lambda\over 6}(\sigma^2+\vec\pi^2)\vec\pi&=&0\nonumber\\
\partial_\mu\partial^\mu\sigma+\mu^2\sigma+{\lambda\over 6}(\sigma^2+\vec\pi^2)\sigma&=&0
\er
These are {\em covariant} under chiral transformations in the sense that the
chiral variation of one of the equations is proportional to the other. When
$\mu^2 > 0$ $\mu$ can be interpreted as the {\em mass} of the chiral multiplet.
Let us now now look at {\em translationally invariant} or {\em vacuum} 
solutions to these equations. They satisfy
\be
\label{vacuum}
(\sigma_c,\vec\pi_c)(\mu^2+{\lambda\over 6}(\sigma_c^2+\vec\pi_c^2))=0
\ee
{\em Vacuum stability} requires $\lambda > 0$. Now if $\mu^2 > 0$ the only
way to satisfy eqn(\ref{vacuum}) is
\be
\sigma_c = 0~~~~~~~~~\vec\pi_c=0
\ee
This is a {\em chirally symmetric} solution in the sense that the chiral
variations of this solution {\em vanish} on this background.

On the other hand if $\mu^2 < 0$ there are non-trivial solutions to eqn(\ref{vacuum})
{\bf Symmetric Solution}\\
\be
\sigma_c = 0~~~~~~~~~\vec\pi_c=0
\ee
{\bf Asymmetric Solution}\\
\be
\label{ssb1}
\sigma_c^2+\vec\pi_c^2 = {6|\mu^2|\over \lambda}
\ee
This condition is in itself {\em an invariant} but the physical vacuum picks 
one of the {\em infinitely} many 
solutions to eqn(\ref{ssb1}). We can use chiral transformations to bring
{\em any solution} to the form $\sigma_c = \sqrt{{6|\mu^2|\over\lambda}}= f_\pi,\vec\pi_c=0$.

The {\em potential} corresponding to the invariant lagrangean in eqn(\ref{fulllag}) can be rearranged as
\be
V(\sigma,\vec\pi) = {\lambda\over 4!}(\sigma^2+\vec\pi^2-f_\pi^2)^2 - {\lambda f_\pi^4\over 4!}
\ee
Now it is clearly seen that the symmetric solution corresponds to $V_{sym}=0$
while the asymmetric solution corresponds to $V_{asym}=-{\lambda f_\pi^4\over 4!}$. Hence the asymmetric solution lies lower and in this case, the true vacuum.

To get the field content of this choice write
\be 
\sigma = f_\pi+\tilde\sigma~~~~~~\vec\pi = \vec{\tilde\pi}
\ee
The {\bf physical fields} defined by {\bf vanishing vacuum values} are now
$(\tilde\sigma,\tilde{\vec\pi})$ fields.

Let us see how this scenario resolves all the five difficulties mentioned
earlier.

i){\bf Nucleon Mass} The interaction term now becomes $g_{NN\pi}f_\pi\bar N N+g_{NN\pi}\bar N \tilde\sigma N$. Not only is a nucleon mass term generated, it
is so done as to satisfy the {\bf Goldberger-Trieman} relation!

iv) The $\sigma\partial_\mu\vec\pi$ term in the axial current now becomes
$f_\pi\partial_\mu\vec\pi+...$ thus resolving this problem.

v) It is easy to see that $\delta_V$ rules in terms of the {\em physical fields}has the same form as before. The axial transformations now become
\be
\delta_A \tilde\sigma = \delta_A \sigma = \vec\omega\cdot\vec\pi;~~~~\delta_A\vec\pi = -f_\pi\vec\omega-\tilde\sigma\vec\omega
\ee
Thus we see that the difficulty with the pion transformation law is also resolved.

Now it will be shown that, rather remarkably, the difficulties ii) and iii)
also get resolved. To see this let us recast eqn(\ref{pieqn}) in terms of the
physical fields
\br
\label{pieqnssb}
\partial_\mu\partial^\mu\vec\pi+\mu^2\vec\pi+{\lambda\over 6}(f_\pi^2+2f_\pi\tilde\sigma+{\tilde\sigma}^2+\vec\pi^2)\vec\pi&=&0\nonumber\\
\partial_\mu\partial^\mu\tilde\sigma+\mu^2f_\pi+\mu^2\tilde\sigma+{\lambda\over 6}(f_\pi^2+2f_\pi\tilde\sigma+{\tilde\sigma}^2+\vec\pi^2)(f_\pi+\tilde\sigma)&=&0
\er
These equations are {\em covariant} wrt the new chiral transformations. On
using the relationship between $\mu,\lambda$ and $f_\pi$ we can recast these
as
\br
\label{ssbeqn}
\partial_\mu\partial^\mu\vec\pi+{\lambda f_\pi\over 3}\tilde\sigma\vec\pi+{\lambda\over 6}(\tilde\sigma^2+\vec\pi^2)&=&0\nonumber\\
\partial_\mu\partial^\mu\tilde\sigma+{\lambda f_\pi^2\over 3}\tilde\sigma+...&=&0
\er
The pion mass term has disappeared and the sigma mass term has turned from
negative mass squared to positive mass squared as it should be for a physical
particle! Thus all the difficulties have been resolved. But now a new physical
particle with a mass $\sqrt{{\lambda f_\pi^2\over 3}}$ is predicted. Does
such a particle exist or not?

While the equations of motion are covariant under the new chiral 
transformation laws, the vacuum solution is {\bf not invariant}. This is called
{\bf Spontaneous breakdown of a symmetry}. General theorems exist to show that 
whenever a {\em continuous symmetry} is broken spontaneously, massless particlesappear in the spectrum of the theory. This is called the {\bf Goldstone Theorem} and the resultant massless particles are called {\bf Goldstone particles}. Thus
{\em pions are the Goldstone bosons of spontaneously broken chiral symmetry}. That they are pseudoscalars is an additional consistency requirement.

One of the consequences of spontaneous symmetry breaking of chiral symmetry
is that additional interactions are produced. For example the ${\lambda\over 4!}(\sigma^2+\vec\pi^2)^2$ term becomes
\br
{\cal L}^{ssb}_{int}&=& {\lambda\over 4!}(f_\pi^2+2f_\pi\tilde\sigma+\tilde\sigma^2+\vec\pi^2)^2\nonumber\\
&=& mass terms+ {\lambda f_\pi\over 3!}{\tilde\sigma}(\tilde\sigma^2+\vec\pi^2)
+{\lambda\over 4!}(\tilde\sigma^2+\vec\pi^2)^2
\er
The additional interaction 
$$
{\lambda f_\pi\over 3!}\tilde\sigma(\tilde\sigma^2+\vec\pi^2)
$$
has many important consequences including the phenomenon of {\bf chiral
cancellations} discussed later. Another important consequence is that
{\em the sigma particle is unstable} with a large width. This is believed to
be one of the reasons for the difficulty in detecting this particle.
\subsection{Finite pion mass}
In real life the pion is not massless but has a small mass and is therefore
referred to as a {\em pseudo-Goldstone boson}. This is not an apt name. So
how does one incorporate this into the picture so far developed? One just
adds an {\bf explicit symmetry breaking term} $-{m_\pi^2\over 2} \vec\pi^2$
to the lagrangean. Actually we would like to add a term that has {\em definite
transformation properties}. So we could add $f_\pi m_\pi^2 \sigma$ instead.
This leads to
\be
\partial_\mu {\vec A}^\mu = f_\pi m_\pi^2 \vec\pi
\ee
which is nothing but the earlier stated PCAC. In the light of modern 
developments like QCD, the symmetry breaking is taken to be $m_q \bar\psi_q \psi_q$ where $q$ refers to the quarks. Consistency between the two pictures then
requires
\be
f_\pi^2m_\pi^2 = m_q <\bar\psi_q \psi_q>
\ee
which is called the {\bf Gellmann-Renner-Oakes relation}.
\newpage
\section{Lecture III: Non Linear Realisation of\\ Chiral Symmetry}
Let us recapitualate the {\em linear realisation} of Chiral Symmetry in terms
of the $(\sigma,\vec\pi)$ fields:
\br
\label{lintr}
\delta_V \sigma = 0~~~~~~~~~~~&~~~~&\delta_V \vec\pi = -{\vec\Lambda}\times \vec\pi\nonumber\\
\delta_A \sigma = \vec\omega\cdot\vec\pi~~~~~~~~~~~&~~~~&\delta_A \vec\pi = -\sigma \vec\omega
\er
As we saw already $\sigma^2+{\vec\pi}^2$ is an {\it invariant}. So it is
possible to obtain a {\em reduced representation} by imposing the invariant
condition
\be
\sigma^2 + {\vec\pi}^2 = f_\pi^2
\ee
The $\sigma$ field is no longer an independent degree of freedom and hence
it can be completely eliminated in terms of the $\vec\pi$ fields 
i.e $\sigma =\sqrt{f_\pi^2-\pi^2}$. This induces a {\em non-linear chiral
transformation} of the pion fields
\be
\label{nonlin1}
\delta_A \vec\pi = -\sqrt{f_\pi^2 - \pi^2}\vec\omega
\ee
In terms of the generators $X_i$ eqn(\ref{nonlin1}) can be recast as
\be
\label{nonlin2}
[X_i,\pi_j] = i\sqrt{f_\pi^2-\pi^2}\delta_{ij}
\ee

The isospin transformations are unchanged and remain {\em linear}. Weinberg
showed that the {\em most general non-linear transformations} are of the
type
\be
\label{nonlin}
[X_i,\pi_j] = i\delta_{ij}f(\pi^2)+\pi_i\pi_j g(\pi^2)
\ee
The functions $f,g$ are not independent but constrained by {\em Jacobi 
Identity}
\be
[X_k,[X_i,\pi_j]]+[\pi_j,[X_k,X_i]]+[X_i,[\pi_j,X_k]]=0
\ee
On using
\be
[X_i,X_j] = i\epsilon_{ijk} T_k;~~~~~~~~~~~~[T_i,\pi_j] = i\epsilon_{ijk}\pi_k
\ee
it is easy to show that
\be
fg-2ff^\prime -2f^\prime g \pi^2 = 1
\ee
It is easy to see that $f$ can not be chosen arbitrarily. For example $f$ can
never vanish. On the other hand $g$ can be chosen arbitrarily and given $g$
we can determine $f$ accordingly. If we take $g=0$ then $f = -\sqrt{c^2-\pi^2}$.
This corresponds to eqn(\ref{nonlin2}) with $c=f_\pi$.

The invariant lagrangean
\be
{\cal L}_{\pi,\sigma} = {1\over 2}[(\partial_\mu\sigma)^2+(\partial_\mu\vec\pi)^2+m^2(\sigma^2+\vec\pi^2)]
\ee
becomes, upon elimination of $\sigma$
\be
{\cal L}_\pi = {1\over 2}[(\partial_\mu \vec\pi)^2 +{\vec\pi\cdot\partial_\mu\vec\pi)^2\over 1-\pi^2}]
\ee
where we have set $f_\pi=1$. This can, following Weinberg, be written in the form
\be
{\cal L}_\pi = {1\over 2}{\cal D}_\mu\vec\pi)^2
\ee
where
\be
{\cal D}_\mu\vec\pi = \partial_\mu\vec\pi+{1\over 2}{1\over \sigma+\sigma^2}\vec\pi \vec\pi\cdot\partial_\mu\vec\pi
\ee
is called the {\em chiral covariant derivative} and is like the covariant
derivatives of {\em general covariant theories}.

Going back to the axial current and the chiral invariant pion-nucleon
Lagrangean, these become
\br
{\vec A}_\mu&=&\sigma\partial_\mu\vec\pi - \vec\pi\partial_\mu\sigma\nonumber\\
&\simeq&\partial_\mu\vec\pi+(\vec\pi \vec\pi\cdot\partial_\mu\vec\pi-\pi^2\partial_\mu\vec\pi)+....
\er
and
\br
{\cal L}_{\pi N} &=& \bar N i\gamma^\mu\partial_\mu N +{1\over 2}({\cal D}_\mu
\vec\pi)^2 -g_{NN\pi}\bar N (\sigma+i\gamma_5\vec\tau\cdot\vec\pi) N\nonumber\\
&=&\bar N(i\gamma^\mu\partial_\mu-M) N +{1\over 2}({\cal D}_\mu\vec\pi)^2 -g_{NN\pi}\bar N({\pi^2\over 2}+i\gamma_5\vec\tau\cdot\vec\pi) N
\er

Another way to understand the non-linear realisation is to imagine studying
the pion-nucleon system at low energies. In the {\em spontaneously broken
phase} $M_\sigma\simeq 5-7 M_\pi$. Thus when all the momenta are small, the
$\sigma$ can considered {\em heavy} and integrated out leaving behind an
effective chiral invariant description of pions and nucleons.

\begin{figure}[htb]
\begin{center}
\mbox{\epsfig{file=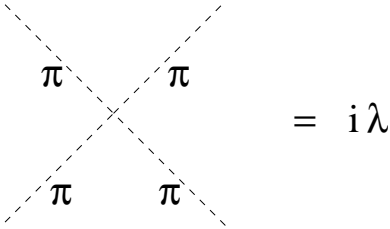,height=3truecm,angle=0}}
\end{center}
\caption{Pi-pi scattering in linear model}
\label{fig:pipilin}
\end{figure}
This brings us to the very important notion of the {\it infrared behaviour
of Goldstone Bosons}. First let us consider the low energy scattering of
pions. The linear theory predicts a constant amplitude which is 
diagrammatically represented in fig(\ref{fig:pipilin}):
\be
A(\pi\pi\rightarrow\pi\pi)\simeq i\lambda
\ee

\begin{figure}[htb]
\begin{center}
\mbox{\epsfig{file=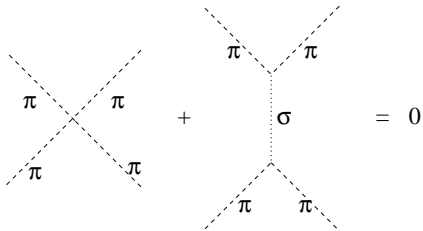,height=3truecm,angle=0}}
\end{center}
\caption{Pi-pi scattering in spontaneously broken phase}
\label{fig:pipissb}
\end{figure}

But the spontaneously broken phase of this theory introduces additional 
processes like $\sigma\rightarrow\pi\pi$ leading to additional contributions 
shown in fig(\ref{fig:pipissb}) which
actually {\em cancel} this contribution to yield a scattering amplitude that
vanishes as $\simeq k^2$!.

\begin{figure}[htb]
\begin{center}
\mbox{\epsfig{file=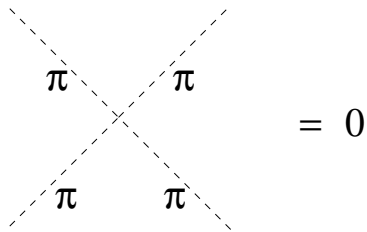,height=3truecm,angle=0}}
\end{center}
\caption{Pi-pi scattering in nonlinear model}
\label{fig:pipinon}
\end{figure}

Same happens in the nonlinear model also. From eqn() the effective interaction 
is given by
\be
{\cal L}_{int} = {1\over 2}(\vec\pi\cdot\partial_\mu\vec\pi)^2
\ee
This means that the pion-pion scattering amplitude, shown in fig(\ref{fig:pipinon}), vanishes $\simeq k^2$!

\begin{figure}[htb]
\begin{center}
\mbox{\epsfig{file=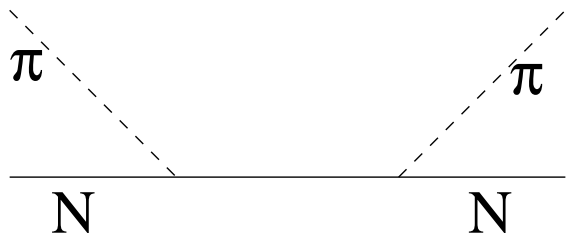,height=3truecm,angle=0}}
\end{center}
\caption{Pion nucleon scattering in linear model}
\label{fig:pinucllin}
\end{figure}

This feature, known as {\em chiral cancellations} is not just restricted
to pion-pion scattering. In fact it is generic to all physical processes
involving {\em Goldstone Bosons}. We illustrate this now for pion-nucleon
scattering. In the linear theory this approaches a constant in the infrared
limit. The relevant amplitude is shown in fig(\ref{fig:pinucllin}).

\begin{figure}[htb]
\begin{center}
\mbox{\epsfig{file=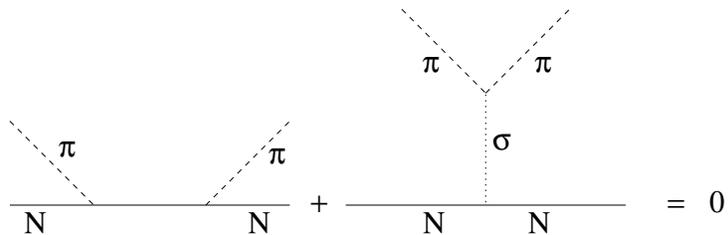,height=3truecm,angle=0}}
\end{center}
\caption{Pion nucleon scattering in spontaneously broken phase}
\label{fig:pinuclssb}
\end{figure}

However in the spontaneously broken phase additional contributions as shown
in fig(\ref{fig:pinuclssb}) exactly cancel this contribution!

The same cancellation takes place in the non-linear model also.

\begin{figure}[htb]
\begin{center}
\mbox{\epsfig{file=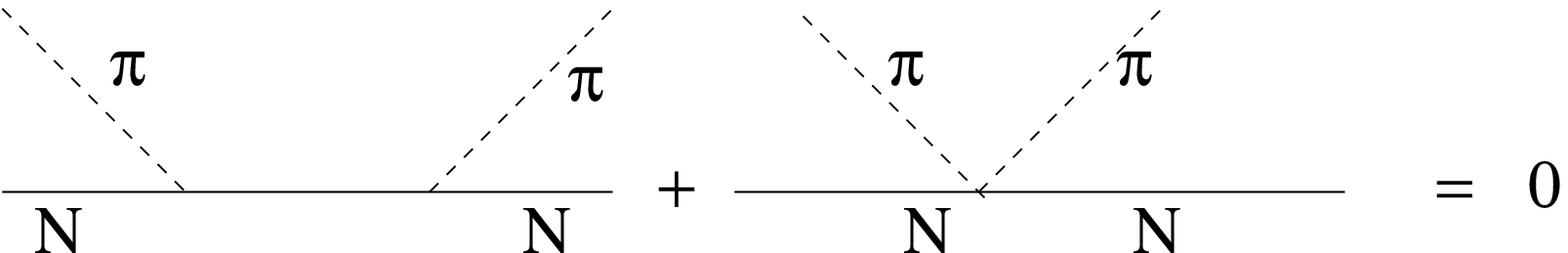,height=3truecm,angle=0}}
\end{center}
\caption{Pion nucleon scattering in nonlinear model}
\label{fig:pinuclnon}
\end{figure}
\subsection{No spontaneous symmetry breaking in the d=4 large N model}
The model we considered can also be thought of as $O(4)$ as $(\sigma,\vec\pi)$
can be arranged into a four-component vector multiplet. In this section we
show that {\em spontaneous symmetry breaking does not happen in d=4} when
we consider $O(N)$ models in the {\em large-N limit}.

Let us start with the definition of the {\em generating functional of connected
Green's functions} $W[{\vec H}]$ of a field theory given by
\be
e^{iW[{\vec H}]} = \int {\cal D}\vec\Phi e^{i(S[\vec\Phi]+{\vec H\cdot\vec\Phi})}
\ee
where
\be
\vec H\cdot\vec\Phi = \int d^4x \vec H(x)\cdot\vec\Phi(x)
\ee
The {\em generating functional of 1PI vertices}, $\gamma[\vec\Phi]$, is 
obtained from $W[\vec H]$ by a {\em Legendre transformation}
\be
\Gamma[\vec \Phi] = W[\vec H] - \vec H\cdot\vec\Phi
\ee
where
\be
\Phi_i(x) = {\delta W[\vec H]\over \delta H_i(x)}
\ee
For the linear $O(N)$-model we have
\be
\label{greengen}
e^{iW[\vec H]} = \int {\cal D}\vec\Phi e^{i\int d^4x [{1\over 2}(\partial_\mu\vec\Phi(x))^2-{U\over 4}(\vec\Phi^2-C^2)^2+\vec H(x)\cdot\vec\Phi(x)]}
\ee
It helps to introduce an {\em auxilliary field} $\chi(x)$ to make the exponent
in eqn(\ref{greengen}) bilinear in $\vec\Phi$:
\be
\label{auxilliary}
e^{iW[\vec H]} = \int {\cal D}\chi{\cal D}\vec\Phi e^{i\int d^4x [{1\over 2}(\partial_\mu\vec\Phi(x))^2-{\chi(x)\over 2}(\vec\Phi^2-C^2)+{1\over 4U}\chi(x)^2+\vec H(x)\cdot\vec\Phi(x)]}
\ee
As $\vec\Phi$ only appears {\em quadratically} it can easily be integrated out
to get
\be
e^{iW[\vec H]} = \int {\cal D}\chi e^{i\int d^4x[{\chi^2(x)\over 4U}+{C^2\over 2}\chi(x)]-{N\over 2}Tr ln \nabla-{i\over 2}\vec H\cdot\nabla^{-1}\cdot\vec H}
\ee
where
\be
\nabla(x,y) = (-\partial^2-\chi(x))\delta (x-y)
\ee
On performing the rescalings
\be
NU = u,C^2=Nc^2,H_i=\sqrt{N}h_i,W[\vec H] = Nw[\vec h]
\ee
we get
\be
e^{iNw[\vec h]} = \int {\cal D}\chi e^{iN
[\int d^4x ({\chi^2\over 4u}+{c^2\over 2}\chi)+{i\over 2}Tr ln \nabla -{1\over 2}\vec h\cdot\nabla^{-1}\cdot\vec h]
}
\ee
Following the standard large-N techniques, the {\em leading order} result is
\be
w_\infty[\vec h] =  
\int d^4x ({\chi^2\over 4u}+{c^2\over 2}\chi)+{i\over 2}Tr ln \nabla -{1\over 2}\vec h\cdot\nabla^{-1}\cdot\vec h
\ee
The {\em classical field} $\phi_i$ where $\vec \Phi = \sqrt{N}\vec \phi$ is
given by
\be
\phi_i(x) = -\nabla^{-1}h_i(x)
\ee
Finally $\Gamma_\infty[\vec \phi,\chi]$ is given by
\be
\Gamma_\infty[\vec \phi,\chi] = \int d^4x[{\chi^2(x)\over 4u}+{c^2\over 2}\chi(x)+{1\over 2}(\partial_\mu\vec\phi(x))^2-{1\over 2}\chi(x)\vec\phi(x)^2]+{i\over 2} Tr ln \nabla
\ee
The resulting {\em effective potential} is
\be
V_{eff}[\vec\phi,\chi] = -{\chi^2\over 4u}-{c^2\over 2}\chi+{1\over 2}\chi\vec\phi^2+{\chi^2\over 64\pi^2}(ln {\chi\over M^2}-{3\over 2})
\ee
The vacua are determined by the stationarity conditions
\be
{\partial V_{eff}\over\partial\chi}=0~~~~~~{\partial V_{eff}\over\partial\phi_a}=0
\ee
That is
\br
\label{Nvacua}
\chi \phi_a&=&0\nonumber\\
-{\chi\over 2u}+{1\over 2}\phi^2-{c^2\over 2}+{1\over 32\pi^2}\chi(ln {\chi\over M^2}-1)&=&0
\er
The spontaneously broken state requires $\chi=0$ and hence $V_{eff}(asy)=0$.
From the second of eqn(\ref{Nvacua}) one sees that this is possible only
if $c^2 > 0$.

On the other hand, the symmetric vacuum condition $\phi^2=0$ and eqn(\ref{Nvacua}) first of all says that $\chi_c\ne 0$ and in fact gives
\be
\label{Nsym}
-{\chi_c\over 2u}-{c^2\over 2}+{1\over 32\pi^2}\chi_c(ln {\chi_c\over M^2}-1)=0
\ee
Using this the effective potential at the symmetric solution can be rewritten
as
\br
V_{eff}(sym) &=& {\chi_c\over 2}(-{\chi_c\over 2u}-{c^2\over 2}+{chi_c\over 32\pi^2}(ln {\chi_c\over M^2}-1))-{c^2\over 4}\chi_c-{\chi_c^2\over 128\pi^2}\nonumber\\
&=& -{c^2\over 4}\chi_c - {\chi_c^2\over 128\pi^2}
\er
now eqn(\ref{Nsym}) says that $\chi_c > 0$ and combining this with the positivity of $c^2$ one sees that $V_{eff}(sym) < 0$ i.e {\bf the symmetric solution
lies below the spontaneously broken solution}!
\end{document}